\documentclass[letterpaper, 10 pt, conference]{ieeeconf}  

\IEEEoverridecommandlockouts
\usepackage[scaled=.8]{beramono} 
\usepackage{microtype}
\usepackage{amsmath} 
\usepackage{amssymb}  
\usepackage[ruled,vlined,linesnumbered]{TemplateFiles/algorithm2e}
\usepackage[usenames]{color}
\usepackage[svgnames]{xcolor}
\usepackage{mathrsfs}
\usepackage{graphicx}
\usepackage{amsfonts}
\usepackage{array}
\usepackage{flafter}
\usepackage{cite}
\usepackage{verbatim}
\usepackage{psfrag}
\usepackage{umoline}
\usepackage{epstopdf}
\usepackage{subcaption} 
\usepackage{hyperref}
\usepackage[capitalize]{cleveref}
\crefformat{equation}{(#2#1#3)}
\Crefformat{equation}{Equation~(#2#1#3)}
\Crefname{equation}{Equation}{Equations}

\hypersetup{%
    pdfborder = {0 0 0}
}
\graphicspath{{./figs/}}

\DeclareMathOperator*{\argmax}{arg\,\!max}

\newcommand{\Cat}{\mathit{Cat}}
\newcommand{\Dir}{\mathit{Dir}}
\newcommand{\Ga}{\mathit{Ga}}

\newtheorem{definition}{Definition}

\title{\LARGE \bf
Semantic-level Decentralized Multi-Robot Decision-Making\\using Probabilistic Macro-Observations
}

\author{Shayegan Omidshafiei$^{1}$, Shih-Yuan Liu$^{1}$, Michael Everett$^{1}$, Brett T. Lopez$^{1}$ \\ Christopher Amato$^{2}$, Miao Liu$^{3}$, Jonathan P. How$^{1}$, John Vian$^{4}$
\thanks{*This work was supported by The Boeing Company.}
\thanks{$^{1}$Laboratory for Information and Decision Systems (LIDS), MIT, Cambridge, MA 02139, USA {\tt\small \{shayegan,syliu,mfe,btlopez,jhow\}@mit.edu}}%
\thanks{$^{2}$College of Computer and Information Science (CCIS), Northeastern University, Boston, MA 02115 USA {\tt\small camato@ccs.neu.edu}}%
\thanks{$^{3}$Thomas J. Watson Research Center, Yorktown Heights, NY 
	 10598, USA {\tt\small miao.liu1@ibm.com }(work completed while the author was at MIT)}%
\thanks{$^{4}$Boeing Research \& Technology, Seattle, WA 98108, USA {\tt\small john.vian@boeing.com}}%
}

\newcounter{Lcount}

\renewcommand{\baselinestretch}{0.97}

\begin{document}
\maketitle
\thispagestyle{empty}
\pagestyle{empty}

\begin{abstract}
Robust environment perception is essential for decision-making on robots operating in complex domains. Intelligent task execution requires principled treatment of uncertainty sources in a robot's observation model. This is important not only for low-level observations (e.g., accelerometer data), but also for high-level observations such as semantic object labels. This paper formalizes the concept of macro-observations in Decentralized Partially Observable Semi-Markov Decision Processes (Dec-POSMDPs), allowing scalable semantic-level multi-robot decision making. A hierarchical Bayesian approach is used to model noise statistics of low-level classifier outputs, while simultaneously allowing sharing of domain noise characteristics between classes. Classification accuracy of the proposed macro-observation scheme, called Hierarchical Bayesian Noise Inference (HBNI), is shown to exceed existing methods. The macro-observation scheme is then integrated into a Dec-POSMDP planner, with hardware experiments running onboard a team of dynamic quadrotors in a challenging domain where noise-agnostic filtering fails. To the best of our knowledge, this is the first demonstration of a real-time, convolutional neural net-based classification framework running fully onboard a team of quadrotors in a multi-robot decision-making domain.
\end{abstract}

\section{Introduction} \label{sec:intro}

\begin{figure}[t]
	\begin{subfigure}[t]{0.23\textwidth}
		\centering
		\includegraphics[height=0.9\textwidth]{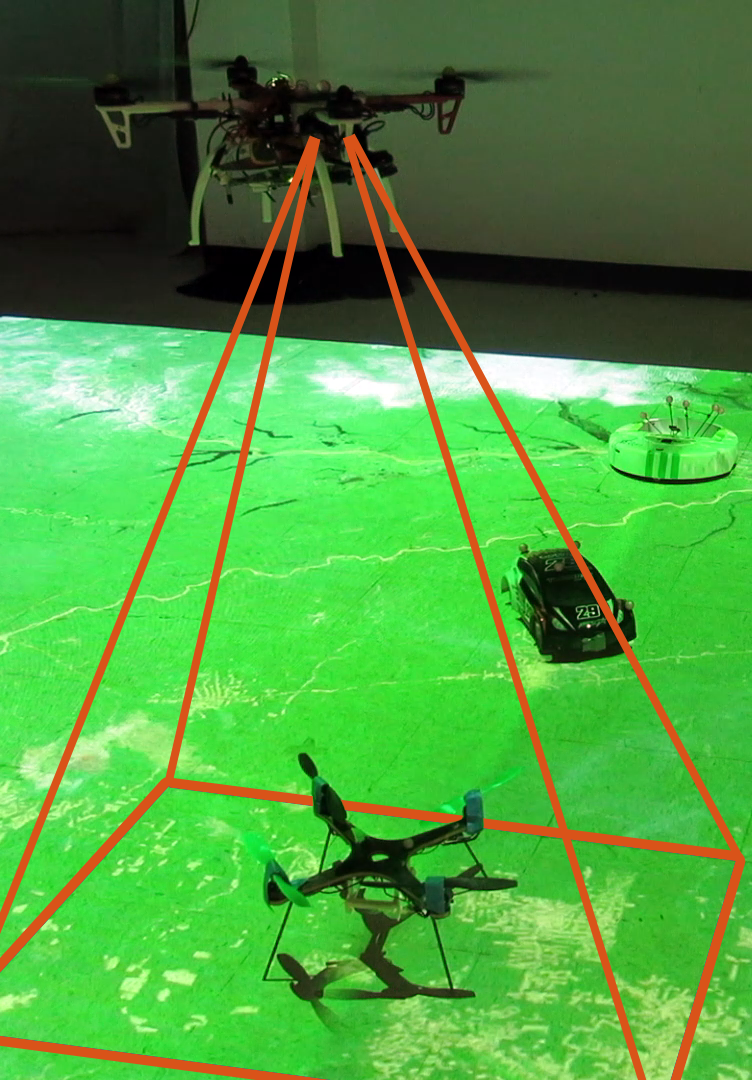}
		\caption{Macro-observations received onboard moving quadrotor.}
		\label{fig:overview_quad_proj}
	\end{subfigure}
	\hfill
	\begin{subfigure}[t]{0.23\textwidth}
		\centering
		\hspace*{-0.2cm}\includegraphics[height=0.9\textwidth]{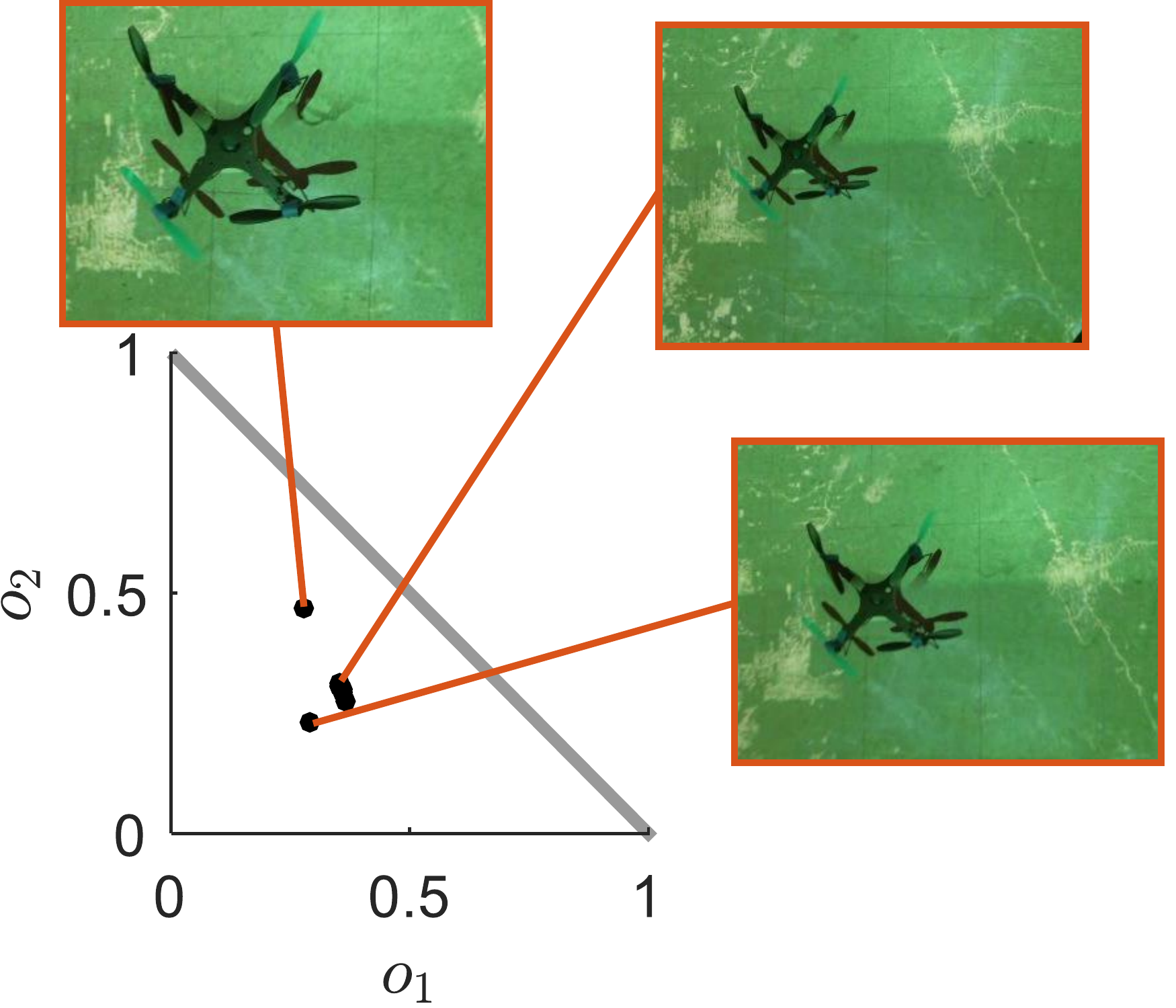}
		\caption{Example classification probability outputs on $\Delta^2$.}
		\label{fig:overview_fig_simplex}
	\end{subfigure}
	\caption{Real-time onboard macro-observations in environments with varying lighting conditions, textures, and motion blur.}
	\label{fig:overview_fig}
\end{figure}

Portable vision sensors, parallelizeable perception algorithms \cite{ILSVRC15}, and general purpose GPU-based computational architectures make simultaneous decision-making and scene understanding in complex domains an increasingly-viable goal in robotics. 
Consider the problem of multi-robot perception-based decision-making in noisy environments, where observations may be low in frame-rate or where semantic labeling is a time-durative process. Each robot may observe an object, infer its underlying class, change its viewpoint, and re-label the object as a different class based on new observations (\cref{fig:overview_fig}). Robots must infer underlying object classes based on histories of past classifications, then use this information to execute tasks in a team-based decision-making setting. 

For autonomous execution of complex missions using perception-based sensors, robots need access to high-level information extending beyond the topological data typically used for navigation tasks. Use of semantic maps (qualitative environment representations) has been recently explored for intelligent task execution \cite{MartinezMozos07b,Galindo:2008:RTP:1453261.1453484,conf/rss/WuLS14}. Yet, limited work has been conducted on semantic-level multi-robot decision-making in stochastic domains. Heuristic labeling rules \cite{Chanel12_ICAPS} or rigid, hand-tuned observation models are failure-prone as they do not infer underlying environment stochasticity for robust decision-making. 
As real-world robot observation processes are notoriously noisy, semantic-level decision-making can benefit from principled consideration of probabilistic observations. 

Cooperative multi-agent decision-making under uncertainty, in its most general form, can be posed as a Decentralized Partially Observable Markov Decision Process (Dec-POMDP) \cite{DecPOMDPBook16}. Yet, infinite horizon Dec-POMDPs are undecidable and finite horizon Dec-POMDPs are NEXP-complete, severely limiting application to real-world robotics \cite{Bernstein02,BernsteinJAIR09}. Recent efforts have improved Dec-POMDP scalability by introducing macro-actions (temporally-extended actions) into the framework, resulting in Decentralized Partially Observable \emph{Semi}-Markov Decision Processes (Dec-POSMDPs) \cite{AmatoRSS15_v2,Omidshafiei15_ICRA,Omidshafiei16_ICRA}. Use of durative macro-actions significantly improves planner scalability by abstracting low-level actions from high-level tasks. 

So far, research focus has been on action-space scalability---no similar work targeting observation-space scalability has been conducted. Further, the scope of the large body of work on Dec-POMDPs has primarily been within the artificial intelligence perspective, with limited focus on traditional robotics applications \cite{DecPOMDPBook16}. While the strength of Dec-POMDPs and Dec-POSMDPs comes from principled treatment of stochasticity, they have primarily been applied to benchmark domains with simple or hand-crafted observation models \cite{DecPOMDPBook16}. Derivation of data-driven, robust observation processes usable for Dec-POSMDP policy search remains a challenge. As planning complexity is exponential in the number of observations, abstraction to meaningful high-level macro-observations (appropriate for the tasks being completed) is desired. Thus, major research gaps exist in leveraging the Dec-POSMDP's full potential for real-world robotics. This paper addresses these issues, providing a high-level abstraction of observation processes and scalability improvements in a similar manner as previous work on macro-actions. 

This paper's primary contribution is a formalization of macro-observation processes within Dec-POSMDPs, with a focus on the ubiquitous perception-based decision-making problem encountered in robotics. A hierarchical Bayesian macro-observation framework is introduced, using statistical modeling of observation noise for probabilistic classification in settings where noise-agnostic methods are shown to fail. The resulting data-driven approach avoids hand-tuning of observation models and produces statistical information necessary for Dec-POSMDP solvers to compute a policy. Hardware results for real-time semantic labeling on a moving quadrotor are presented, with accurate inference in settings with high perception noise. The entire processing pipeline is executed onboard a quadrotor at approximately 20 frames per second. The macro-observation process is then integrated into a Dec-POSMDP planner, with demonstration of semantic-level decision-making executed on a quadrotor team performing a perception-based health-aware disaster relief mission.

\section{Decentralized Multi-Robot Decision-Making} \label{sec:decposmdps}
This section summarizes the Dec-POSMDP framework, a decentralized decision-making process targeting large-scale multi-agent problems in stochastic domains. The Dec-POSMDP addresses scalability issues of Dec-POMDPs by incorporating belief-space macro actions, or temporally-extended actions. For details on Dec-POSMDP fundamentals, we refer readers to our previous work \cite{Omidshafiei16_ICRA,AmatoRSS15_v2,Omidshafiei15_ICRA}.

Robots involved in Dec-POSMDPs operate in belief space, the space of probability distributions over states, as they only perceive noisy observations of the underlying state. Solving a Dec-POSMDP results in a hierarchical decision-making policy, where a macro-action (MA) $\pi^{(i)} \in \mathbb{T}^{(i)}$ is first selected by each robot $i \in \mathbb{I}$, and low-level (primitive) actions are conducted within the MA until an $ \epsilon $-neighborhood of the MA's belief milestone $\check{b}^{goal}$ is reached.\footnote{We denote a generic parameter $p$ of the $i$-th robot as $p^{(i)}$, joint parameter of the team as $\bar{p}$, and joint parameter at timestep $k$ as $\bar{p}_{k}$.} This neighborhood defines a \emph{goal belief node} for the MA, denoted $ B^{goal}\!=\!\{b\!:\!\|b-\check{b}^{goal}\|\!\!\leq\!\epsilon \} $. Each MA encapsulate a low-level POMDP involving primitive actions $u_{t}^{(i)}$ and observations $o_{t}^{(i)}$.

\begin{definition}
	The Dec-POSMDP is defined below:
\end{definition}
\begin{itemize}
	\item $\mathbb{I}=\{1,2,\ldots,n\}$ is the set of heterogeneous robots.
	\item $\mathbb{B}^{(1)}\times\mathbb{B}^{(2)}\times\ldots\times\mathbb{B}^{(n)}\times\mathbb{X}^{e} $ is the underlying belief space, where $ \mathbb{B}^{(i)} $ is the set of belief milestones of the $ i $-th robot's MAs and $\mathbb{X}^{e}$ is the environment state space.
	\item $ \bar{\mathbb{T}}=\mathbb{T}^{(1)}\times\mathbb{T}^{(2)}\ldots\times\mathbb{T}^{(n)} $ is joint independent MA space, where $ \mathbb{T}^{(i)} $ is the set of MAs for the $ i $-th robot. $\bar{\pi} = \{\pi^{(1)},\ldots,\pi^{(n)}\}$ is the team's joint MA.
	\item $ \bar{\breve{\mathbb{O}}}^{e}=\{\bar{\breve{o}}^{e} \} $ is the set of all joint MA-observations. 
	\item $ P(\bar{b}',x^{e'},k|\bar{b},x^{e};\bar{\pi}) $ is the high-level transition probability model under MAs $ \bar{\pi} $ from $ (\bar{b},x^{e}) $ to $ (\bar{b}',x^{e'}) $.
	\item $ \bar{R}^{\tau}\!(\bar{b},x^{e};\bar{\pi}) $ is the generalized reward of taking a joint MA $ \bar{\pi} $ at $(\bar{b},x^{e})$, where $\bar{b}$ is the joint belief.
	\item $ P(\bar{\breve{o}}^{e}|\bar{b},x^{e}) $ is the joint observation likelihood model, with observation $\bar{\breve{o}}^{e}=\{\breve{o}^{e(1)},\breve{o}^{e(2)},\ldots,\breve{o}^{e(n)} \} $.
	\item $\gamma \in [0, 1)$ is the reward discount factor.
\end{itemize}

Let $\mathbb{X}^{e}$ be the high-level or macro-environment state space, a finite set describing the state space extraneous to robot states (e.g., an object in the domain). An observation of the \emph{macro-environment state} $x^{e} \in \mathbb{X}^{e}$ is denoted as the \emph{macro-observation} $o^{e(i)}$. Upon completion of its MA, each robot makes macro-observation $o^{e(i)}$ and calculates its final belief state, $b^{f(i)}$. This macro-observation and final belief are jointly denoted as $\breve{o}^{e(i)} = (o^{e(i)}, b^{f(i)})$.

The history of executed MAs and received high-level observations is denoted as the \emph{MA-history},
\begin{align}
\xi^{(i)}_{k}=\{\breve{o}^{e(i)}_{0},\pi^{(i)}_{0},\breve{o}^{e(i)}_{1},\pi^{(i)}_{1},\ldots,\breve{o}^{e(i)}_{k-1},\pi^{(i)}_{k-1},\breve{o}^{e(i)}_{k}\}.
\end{align}

The transition probability $P(\bar{b}',x^{e'},k|\bar{b},x^{e};\bar{\pi})$ from $(\bar{b},x^{e})$ to $(\bar{b}',x^{e'})$ under joint MA $\bar{\pi}$ in $k$ timesteps is \cite{Omidshafiei16_ICRA},
\begin{align} \label{eq:MA_trans_prob_epoch}
& P(\bar{b}',x^{e'},k|\bar{b}_{0},x^{e}_{0},o^{e}_{k};\bar{\pi})
= P(x^{e}_k,\bar{b}_k|\bar{b}_{0},x^{e}_{0},o^{e}_{k};\bar{\pi})\nonumber\\
&~~~~~~=\sum_{x^{e}_{k-1},\bar{b}_{k-1}}\Big[P(x^{e}_k|x^{e}_{k-1},o^{e}_{k};\bar{\pi}(\bar{b}_{k-1}))\times \\
&P(\bar{b}_k|x^{e}_{k-1},\bar{b}_{k-1};\bar{\pi}(\bar{b}_{k-1})) P(x^{e}_{k-1},\bar{b}_{k-1}|x^{e}_{0},\bar{b}_{0};\bar{\pi}(\bar{b}_{0}))\Big].\nonumber
\end{align}

The generalized team reward for a discrete-time Dec-POSMDP during execution of joint MA $\bar{\pi}$ is defined \cite{Omidshafiei16_ICRA},
\begin{align}\label{eq:gen_reward_k=0}
\!\!\!\!\!\bar{R}^{\tau}\!(\bar{b},x^{e};\bar{\pi}) \!=\! \mathbb{E}\!\left[\!\sum_{t=0}^{\tau-1}\!\!\gamma^{t}\!\bar{R}(\bar{x}_{t},x^{e}_{t}\!,\bar{u}_{t})|P(\bar{x}_{0})\!=\!\bar{b},x^{e}_{0}\!=\!x^{e}\!;\bar{\pi}\!\right]
\end{align}
where $\tau = \min_{i}\min_{t}\{t:b^{(i)}_{t}\in B^{(i),goal} \}$ is the timestep at which robot $i$ completes its current MA, $\pi^{(i)}$. Note that $\tau$ is itself a random variable, since MA completion times are also non-deterministic. Thus, the expectation in \eqref{eq:gen_reward_k=0} is taken over MA completion durations as well. In practice, sampling-based approaches are used to estimate this expectation.

MA selection is dictated by the \emph{joint high-level policy}, $\bar{\phi} = \{\phi^{(1)}, \ldots, \phi^{(n)} \}$. Each robot's high-level policy $\phi^{(i)}$ maps its MA-history $\xi^{(i)}_{k}$ to a subsequent MA $\pi^{(i)}$ to be executed. The joint value under policy $\bar{\phi}$ is,
\begin{align}\label{eq:decposmdp_eval}
&\!\bar{V}^{\bar{\phi}}(\bar{b},x^{e})
=\mathbb{E}\left[\sum_{k=0}^{\infty}\gamma^{t_{k}}\bar{R}^{\tau}(\bar{b}_{t_{k}},x^{e}_{t_{k}};\bar{\pi}_{t_{k}})|\bar{b}_{0},x^{e}_{0};\bar{\phi}\right]\\
&\!\!=\bar{R}^{\tau}\!(\bar{b},x^{e};\bar{\pi}) \! \notag + \!\!
\sum_{k=1}^\infty\gamma^{t_k}\!\!\!\!\!\!\sum_{\bar{b}',x^{e'},\bar{o}^{e'}}\!\!\!\!\!\!P(\bar{b}'\!,x^{e'}\!\!,\bar{o}^{e'}\!\!,k|\bar{b},x^{e};\bar{\pi})\bar{V}^{\bar{\phi}}(\bar{b}'\!,x^{e'}).
\end{align}
The optimal joint high-level policy is then,
\begin{align}\label{eq:decposmdp_problem}
\bar{\phi}^{*}=\argmax_{\bar{\phi}}\bar{V}^{\bar{\phi}}(\bar{b},x^{e}).
\end{align}

To summarize, the Dec-POSMDP is a hierarchical decision-making process which involves finding a joint high-level policy $\bar{\phi}$ dictating the MA $\pi^{(i)}$ each robot $i \in \mathbb{I}$ conducts based on its history of executed MAs and received high-level observations. Within each MA, the robot executes low-level actions $u_t^{(i)}$ and perceives low-level observations $o_t^{(i)}$. Therefore, the Dec-POSMDP is an abstraction of the Dec-POMDP process which treats the problem at the high macro-action level to significantly increase planning scalability. 

\section{Semantic Macro-Observations} \label{sec:macro_observations}
This section formalizes Dec-POSMDP macro-observations. It also outlines the sequential-observation classification problem for macro-observation models and introduces a hierarchical Bayesian scheme for semantic-level macro-observations.

\subsection{Macro-Observation Processes} \label{sec:macro_observation_processes}
Dec-POSMDPs naturally embed state and macro-action uncertainty into a high-level decision-making process. In a similar manner, task planning can benefit from the robot's high-level understanding of the environment state. Previous research has focused on formal definitions of MAs in terms of low-level POMDPs and on algorithms for automatically generating them \cite{Omidshafiei15_ICRA}. Yet, no formal work on automatic macro-observation generation has been done to date. Benchmark domains used to test Dec-POSMDP search algorithms use simplistic or hand-coded high-level observation processes, which are subsequently sampled during the evaluation phase of policy search algorithms \cite{Omidshafiei16_ICRA,Omidshafiei15_ICRA}. In contrast, this paper provides a foundation for deriving meaningful, data-driven macro-observations. We formally define macro-observations herein by distinguishing them from low-level observations: 
\begin{definition}\label{def:macro_observations}
	Macro-observations are durative, generative probabilistic processes within which sequences of low-level observations are filtered, resulting in a semantic-level observation of the environment.
\end{definition}

Macro-observations allow each robot's noisy semantic perception of the world to affect its task selection. Just as MAs provide an abstraction of low-level actions to a high-level task (e.g., ``Open the door"), macro-observations abstract low-level observations to a high-level meaningful understanding of the environment state (e.g., ``Am I in an office?"). 

For uncertainty-aware planning, Dec-POSMDP policy search algorithms require sampling of the domain transition and observation model distributions discussed in \cref{sec:decposmdps}. Thus, the following distributions must be calculable for any robot's derived macro-observation process:
\begin{enumerate}
	\item a semantic output distribution $ P(o^{e}|b^{(i)},x^{e}) $ of underlying macro (environment) state
	\item a distribution over computation time $\tau$ 
\end{enumerate}

While low-level observation processes can be treated as instantaneous for simplicity, observations related to scene semantics require non-negligible computation time which must be accounted for in the planner. Dec-POSMDPs seamlessly take this computation time into account. \cref{def:macro_observations} provides a natural representation for real-world high-level observation processes, as they are durative (i.e., take multiple timesteps to process low-level data). Further, this computation time is non-deterministic (e.g., the amount of time needed to answer ``Am I in an office?" is conditioned on scene lighting). The existing Dec-POSMDP transition dynamics in \eqref{eq:gen_reward_k=0} take an expectation over MA completion times. As every macro-observation is perceived following an MA, the time distribution in \eqref{eq:gen_reward_k=0} can seamlessly include macro-observation computation time.

The result is a particularly powerful semantic-level decision-making framework, as MAs targeting desired macro-observations can be embedded in the Dec-POSMDP (e.g., ``Track object until its class is inferred with 95\% confidence"). The next sections focus on development of an automatic process which provides Dec-POSMDP solvers with the two necessary macro-observation distributions (semantic output distribution and computation time distribution).


\subsection{Sequential Classification Filtering Problem}\label{subsec:classification_filtering_problem}

We now detail generation of macro-observations in the context of probabilistic object classification. Specifically, consider a ubiquitous decision-making scenario where a robot observes a sequence of low-level classifier outputs and must determine its surrounding environment state or class of an object in order to choose a subsequent task to execute. A unique trait of robotic platforms is locomotion, allowing observations of an object or scene from a variety of viewpoints (\cref{fig:overview_fig}). This motivates the need for a sequential macro-observation process using the history of classification observations made by the robot throughout its mission. In contrast to na{\"i}ve reliance on frame-by-frame observations, sequential filtering offers increased robustness against domain uncertainty (e.g., camera noise, lighting conditions, or occlusion).

\begin{figure}[t!]
	\centering
	\begin{subfigure}[t]{0.11\textwidth}
		\centering
		\includegraphics[trim={0.3cm 0.1cm 0.5cm 0},width=1\textwidth]{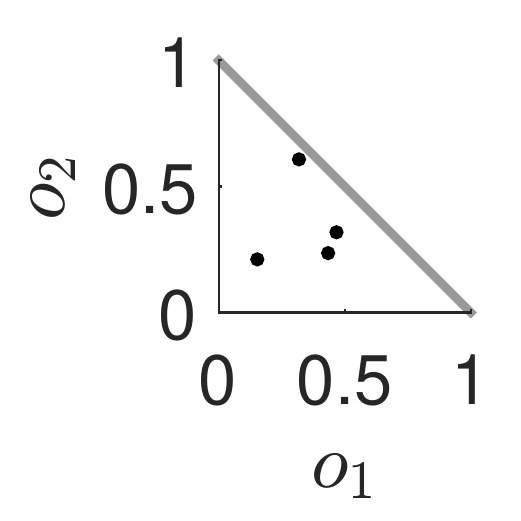}
		\caption{4 low-level classifier observations.}
		\label{fig:obs_model_example}
	\end{subfigure}
	\hfill
	\begin{subfigure}[t]{0.11\textwidth}
		\centering
		\includegraphics[trim={0.5cm 0.1cm 0.5cm 0},width=1\textwidth]{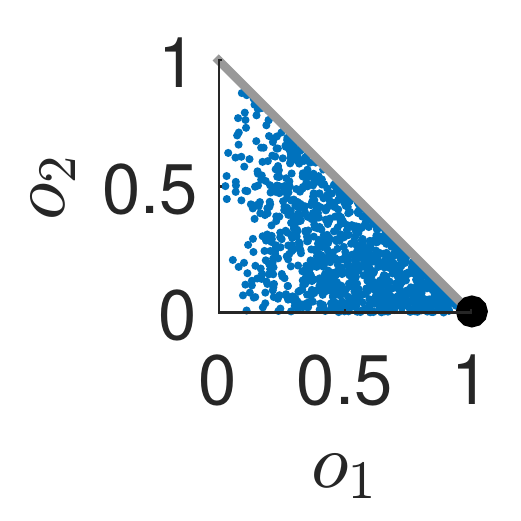}
		\caption{Classifications for $c=1$, $\theta_{c=1} = 1$.}
		\label{fig:obs_model_class_1}
	\end{subfigure}
	\hfill
	\begin{subfigure}[t]{0.11\textwidth}
		\centering
		\includegraphics[trim={0.5cm 0.1cm 0.5cm 0},width=1\textwidth]{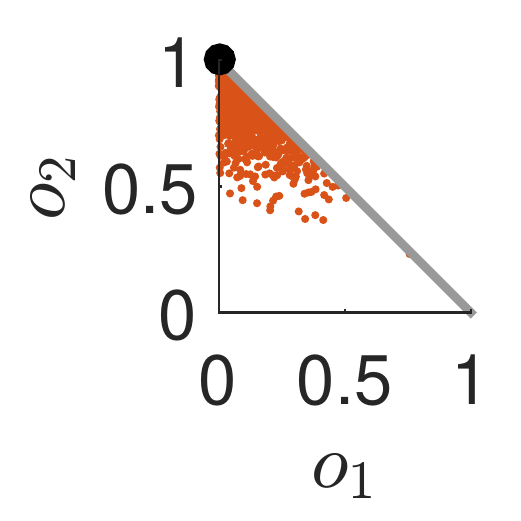}
		\caption{Classifications for $c=2$, $\theta_{c=2} = 6$.}
		\label{fig:obs_model_class_2}
	\end{subfigure}
	\hfill
	\begin{subfigure}[t]{0.11\textwidth}
		\centering
		\includegraphics[trim={0.5cm 0.1cm 0.5cm 0},width=1\textwidth]{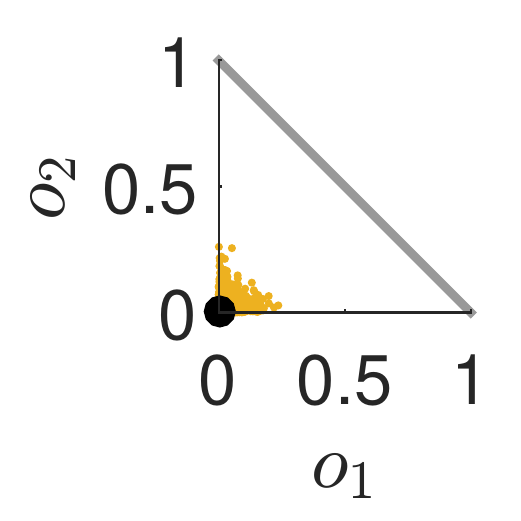}
		\caption{Classifications for $c=3$, $\theta_{c=3} = 20$.}
		\label{fig:obs_model_class_3}
	\end{subfigure}
	\caption{Motivating macro-observation example with 3 classes. Each point represents a single low-level observation $o_i \in \Delta^2$.}
	\label{fig:example_observation_models}
\end{figure}

In settings with high observation noise, or where training data is not representative of mission data, statistical analysis of low-level classifier outputs both improves accuracy of macro-observations and provides useful measures of perception uncertainty. As a motivating example, consider the 3-class scenario in \cref{fig:example_observation_models}. A low-level classifier predicts the probability of a single image belonging to each class. A sequence of images results in a corresponding sequence of observed class probabilities, as in \cref{fig:obs_model_example} for a 4-image sequence. This makes inference of the underlying class nontrivial. 


Let us formalize the problem of constructing semantic macro-observations using streaming classification outputs. Given input feature vector $f_i$ at time $i$, an M-class probabilistic classifier outputs low-level probability observation $o_i= (\!o_{i,1},\ldots,o_{i,m},\ldots, o_{i,M}\!)$, where $o_{i,m}$ is the raw probability of $f_i$ belonging to the $m$-th class (e.g., \cref{fig:obs_model_example}). Thus, $o_i$ is a member of the $(M-1)$-simplex, $\Delta^{M-1}$.


In object classification, $f_i$ may be an image or feature representation thereof, and $o_{i,m}$ represents probability of the object belonging to the $m$-th class. This probabilistic classification is conducted over a sequence of $N$ images $f_{1:N}$, resulting in a stream of class probability observations $o_{1:N}$. In robotics, this macro-observation process is inherently durative as multiple low-level observations of the object need to be perceived to counter domain noise. 
Simply labeling the object as belonging to the class with maximal probability, $\argmax_{m} (o_{i,m})$, can lead to highly sporadic outputs as the image sequence progresses. A filtering scheme using the history of classifications $o_{1:N}$ is desired, along with the two aforementioned characterizing macro-observation distributions necessary for Dec-POSMDP search algorithms.

Prior work on aggregation of \emph{multiple} classifiers' predictions can be extended to single-classifier multi-observation filtering, where in each case, the posterior outputs would become macro-observation $o^e$. Fixed classifier combination rules offer simplicity in implementation at the cost of sub-optimality. One example is the max-of-mean approach \cite{journals/tsmc/XuKS92}, where the $m$-th class posterior probability, $o'_{m}$, is the mean of observed probabilities throughout the image sequence,
\begin{align}
o'_{m} = \frac{1}{N} \sum_{i=1}^{N}o_{i,m}\,\,\,\, \forall m \in \{1,\ldots,M\}.
\end{align}

Another strategy is voting-based consensus \cite{Florian:2002:MCC:1118693.1118697}, with posterior class $c$ chosen based on the highest number of votes from all individual prediction observations $o_i$,
\begin{align}
c = \argmax_{c' \in \{1,\ldots,M\}} \sum_{i}\delta(c',\argmax_{m \in \{1,\ldots,M\}} o_{i,m})
\end{align}
where $\delta(\cdot,\cdot)$ is the Kronecker delta function. 

The above approaches do not exploit the probabilistic nature of underlying classifier outputs, $o_i$. A Bayes filter offers a more principled treatment of the problem. For example, binary Bayes filters are a popular approach for occupancy grid filtering and object detection \cite{thrun2001,conf/icra/CoatesN10}, where repeated observations are filtered to determine occupancy probability or presence of an object (both are $M=2$ class cases, with classes `occupied/present' or `empty/absent'). 
Binary Bayes filters can be extended to M-class recursive classification by applying Bayes rule and a Markovian observation assumption,
\begin{align}
\!\!\!\!\!P(c\!=\!m|f_{1:N})\! \propto & \frac{P(c = m|f_N)}{P(c = m)}P(c = m|f_{1:N-1}),\label{eq:bayes_filter_3}
\end{align}
where $P(c=m)$ is the prior class distribution and $P(c=m|f_N) = o_{N,m}$. This Bayes filter assumes a fixed underlying class, henceforth called a Static State Bayes Filter (SSBF).

Though SSBF allows probabilistic filtering of classifier outputs, it assigns equal confidence to each observation $o_i$ in its update. It takes equal amount of evidence for a class to ``cancel out" evidence against it, an issue encountered in Bayes-based occupancy mapping \cite{conf/fsr/YguelAL07}. In settings with heterogeneous classifier performance, this approach performs poorly. One class may be particularly difficult to infer in a given domain, increasing probability of misclassifications compared to other classes. In our motivating example, \cref{fig:obs_model_class_1,fig:obs_model_class_2,fig:obs_model_class_3} illustrate noisy classification samples for the 3 underlying object classes. Class $c=1$ (\cref{fig:obs_model_class_1}) is particularly difficult to classify, with a near-uniform distribution of $o_i$ throughout the simplex, in contrast to high-accuracy classifications of $c=3$ (\cref{fig:obs_model_class_3}). In this case, given uniform observations throughout the simplex \emph{and} knowledge of underlying classifier noise, the filter update weight on underlying class $c=1$ should be higher than $c=3$, since the classifier outputs are most sporadic for class $c=1$. 

The critical drawback of the above approaches is that they simply \emph{filter}, but do not \emph{model}, the underlying observation process.  As discussed earlier in \cref{sec:macro_observation_processes}, generative high-accuracy macro-observation models are necessary for Dec-POSMDP policy search algorithms \cite{Omidshafiei16_ICRA,AmatoRSS15_v2}. Perception-based observations are highly complex and involve images/video sequences generated from the domain, making them (currently) impossible to replicate in these offline search algorithms. While it may be tempting to use hand-coded generative distributions for the above \emph{filter-based} macro-observation processes during policy search, such an approach fails to exploit the primary benefit of POMDP-based frameworks: the use of data-driven noise models which result in policies that are robust in the real world.


\subsection{Hierarchical Approach for Semantic Macro-Observations}\label{subsec:hbni}
This section introduces a generative macro-observation model titled Hierarchical Bayesian Noise Inference (HBNI), which infers inherent heterogeneous classifier noise. HBNI provides a compact, accurate, generative perception-based observation model, which is subsequently used to sample the two macro-observation distributions in Dec-POSMDP solvers. The combination of Dec-POSMDPs with HBNI macro-observations allows robust, probabilistic semantic-level decision-making in settings with limited, noisy observations.

To ensure robustness against misclassifications, HBNI involves both noise modeling and classification filtering, making it a multi-level inference approach. Given a collection of image class probability observations $o_{1:N}$ (\cref{fig:obs_model_example}), the underlying class for each image $f_i$ is inferred while modeling classifier noise distributions.
\begin{figure}[t]
	\centering
	\includegraphics[trim={0 0.3cm 0 -0.4cm},width=0.75\linewidth]{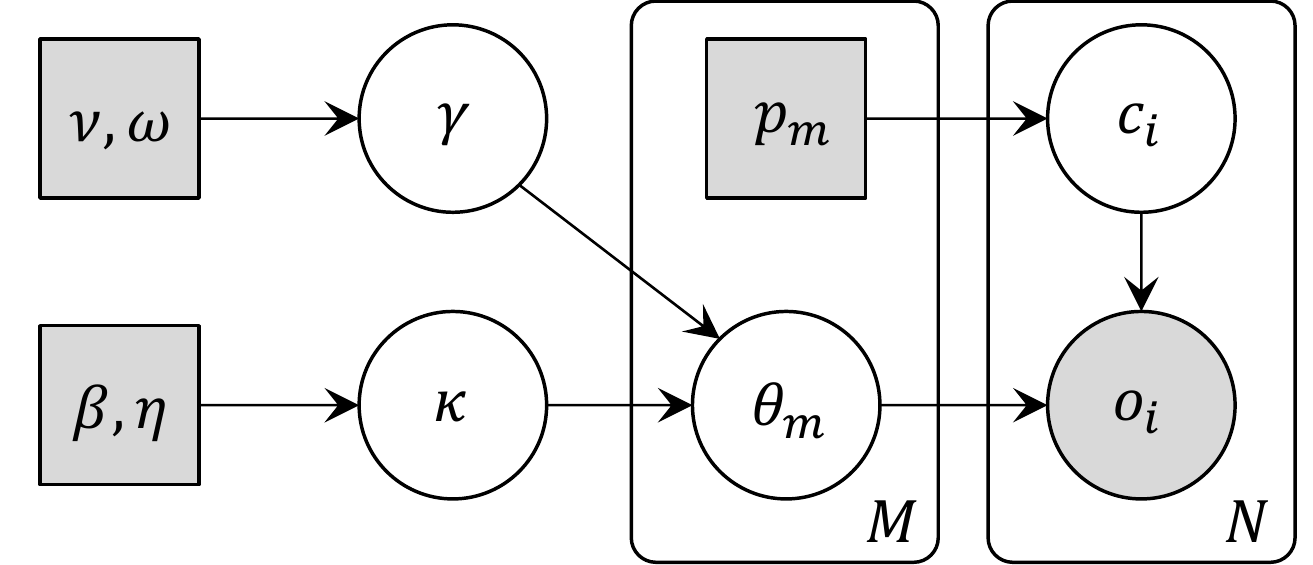}
	\caption{The HBNI model, with per-class noise parameters $\theta_m$ and shared hyperparameters, $\kappa, \gamma$.}\label{fig:unique_obs_hyperprior_bayes_plate_v3}
\end{figure}

Hierarchical Bayesian models allow multi-level abstraction of uncertainty sources \cite{good80}. This is especially beneficial in stochastic settings targeted by Dec-POSMDPs, which have layered sources of uncertainty. In semantic labeling, for instance, parameterization of the classifier confidence for the $M$ classes can be modeled using a set of noise parameters $\theta_{1:M}$. Moreover, it is beneficial to model the relationship between noise parameters through a shared prior (\cref{fig:unique_obs_hyperprior_bayes_plate_v3}). Consider, for instance, a robot performing object classification using a low-quality camera or in a domain with poor visibility. In this setting, observations may be noisier than expected a priori, indicating presence of a high-level, class-independent uncertainty source. This information should be shared amongst all class models, allowing more accurate modeling of domain uncertainty through the noise parameters. Layered sharing of statistical information between related parameters is a strength of hierarchical Bayesian models, and has been demonstrated to increase robustness in posterior inference compared to non-hierarchical counterparts \cite{conf/icml/HugginsT15}. 

\cref{fig:unique_obs_hyperprior_bayes_plate_v3} illustrates the graphical model of HBNI. A categorical prior is used for classes, 
\begin{align}\label{eq:cat_prior}
c_{i} \sim \Cat(p_{1:M}),
\end{align}
where $p_i \in \Delta^{M-1} \,\, \forall i \in \{1,\ldots,M\}$. This allows integration of prior domain knowledge into HBNI. A Dirichlet observation model is used for raw classifier outputs $o_i \in \Delta^{M-1}$, 
\begin{align}
o_i \sim \Dir(\theta_{c_{i}}\vec{1}_{c_{i}}+\vec{1}),
\end{align}
where $\theta_{c_i} \geq 0$ is a scalar noise parameter for the associated class, $\vec{1}_{c_{i}}$ is an $M\times1$ categorical vector with the $c_i$-th element equal to 1 and remaining element equal to zero, and $\vec{1}$ is an $M\times1$ vector of ones. Each class observation $o_i$ has an associated class label $c_i$, which in turn links $o_i$ to the appropriate noise parameter $\theta_{c_i}$ (the $c_i$-th element of parameter set $\{\theta_1,\ldots,\theta_M\}$). This choice of parameterization offers two advantages. First, the selection of $\theta_{c_i}$ provides a direct, intuitive measure of noise for the classifier observations. As in \cref{fig:obs_model_class_1,fig:obs_model_class_2,fig:obs_model_class_3}, $\theta_{c_i}$ is the Dirichlet concentration parameter and is related to the variance of the classification distribution. Low values of $\theta_{c_i}$ imply high levels of observation noise, and vice versa. A second advantage is that it simplifies the posterior probability calculations used within Markov chain Monte Carlo (MCMC) inference, as discussed below.

A gamma prior is used for noise parameter $\theta_m$,
\begin{align}
\theta_m \sim \Ga(\kappa,\gamma),
\end{align}
where $\kappa$ and $\gamma$ themselves are treated as unknown hyperparameters. The role of $\kappa$ and $\gamma$ is to capture high-level sources of domain uncertainty, allowing sharing of cross-class noise statistics. Gamma priors (parameterized by ($\beta, \eta)$ and $(\nu,\omega)$) were also used for these hyperparameters in our experiments, although results showed low sensitivity to this prior choice.

Given raw class probability observations $o_{1:N}$, the posterior probability of noise parameters and associated classes is,
\begin{align}
\!\!\!P&(\theta_{1:M},c_{1:N},\kappa,\gamma|o_{1:N}) \nonumber \\
&\propto \prod_{i=1}^{N} P(o_i|\theta_{c_i}, c_i)P(c_i) \prod_{m=1}^{M}P(\theta_m|\kappa,\gamma)P(\kappa)P(\gamma)\\
&= \prod_{i=1}^{N} \Big[\Dir(o_i;\theta_{c_i}\vec{1}_{c_i}+\vec{1})\Cat(c_i;p_{1:M})\Big]\nonumber\\ 
& \phantom{{} =} \times \prod_{m=1}^{M}\Ga(\theta_m;\kappa,\gamma) \Ga(\kappa;\beta,\eta)\Ga(\gamma;\nu,\omega).\label{eq:theta_c_inference_orig}
\end{align}
This allows inference of noise parameters $\theta_{1:M}$ and hyperparameters $\kappa$ and $\gamma$ using the collection of observed data $o_{1:N}$. The computational complexity of \cref{eq:theta_c_inference_orig} can be further reduced. The log of the prior \cref{eq:cat_prior} is simply $\log \Cat(c_i;p_{1:M}) = \log p_{c_i}$. To efficiently compute $\log\Dir(o_i;\theta_{c_i}\vec{1}_{c_i}+\vec{1})$, consider a notation change. Letting $\bar{\alpha} = \{\alpha_1,\ldots,\alpha_M\}=\theta_{c_i}\vec{1}_{c_i}+\vec{1}$,
\begin{align}
\Dir(o_i;\bar{\alpha})=\frac{1}{B(\bar{\alpha})} \prod_{m=1}^{M}o_{i,m}^{\alpha_m-1},\label{eq:dirich_alpha_bar}
\end{align}
with $B(\cdot)$ as the Beta function. Based on the definition of $\bar{\alpha}$,
\begin{align}
\alpha_m-1=\begin{cases}
\theta_{c_i}, & m=c_i.\\
0, & \text{$m \neq c_i$}.
\end{cases} \label{eq:alpha_cases}
\end{align}
Combining \cref{eq:alpha_cases} with \cref{eq:dirich_alpha_bar} and taking the log,
\begin{align}
\!\!\!\!\!\!\log&\Dir(o_i;\bar{\alpha})\nonumber\\
=&-\log B(\bar{\alpha}) + \theta_{c_i}\log o_{i,c_i}\\
=&-\sum_{m=1}^{M}\log\Gamma(\alpha_m)+\log\Gamma(\sum_{m=1}^{M}\alpha_m) + \theta_{c_i}\!\log o_{i,c_i},\label{eq:log_dir_beta_expanded}
\end{align}
where $\Gamma$ is the gamma function. Note that as per \cref{eq:alpha_cases}, 
\begin{align}
\log\Gamma(\alpha_m)=\begin{cases}
\log\Gamma(1+\theta_{c_i}), & m=c_i.\\
0, & \text{$m \neq c_i$}.
\end{cases} \label{eq:log_gamma_cases}
\end{align}
and $\sum_m \alpha_m = M + \theta_{c_i}$. Thus, the Dirichlet log-posterior is,
\begin{align}
\!\!\!\log\Dir&(o_i;\bar{\alpha})\nonumber\\	=\!-&\!\log\Gamma(1+\theta_{c_i})+\log\Gamma(M+\theta_{c_i}) +\theta_{c_i}\!\log o_{i,c_i}.\label{eq:log_dir_beta_final}
\end{align}
Finally, the log-probability of $\theta_m$ (and similarly $\kappa$, $\gamma$) is,
\begin{align}
\log\Ga(\theta_m;\kappa,\gamma) \propto (\kappa-1)\log\theta_m -\theta_m\gamma^{-1}.\label{eq:gamma_log_final}
\end{align}
To summarize, the log of \cref{eq:theta_c_inference_orig} is efficiently computed by combining \eqref{eq:log_dir_beta_final} and \eqref{eq:gamma_log_final}. An MCMC approach is used to calculate the posterior distribution over noise parameters ($\theta_{1:M}$) and hyperparameters ($\kappa$, $\gamma$). This allows a history of observations $o_{1:N}$ to be filtered using the noise distributions, resulting in posterior class probabilities,
\begin{align}
P&(c=m|o_{1:N},\theta_{1:M}) \nonumber\\
&\propto P(o_{1:N}|\theta_m,c=m)P(c=m)\\
&=P(o_N|\theta_m,c=m)\Big[\prod_{i=1}^{N-1}P(o_i|\theta_m,c=m)p_m\Big],\label{eq:classification_update_original}
\end{align}
where $c$ is conditionally independent of $\kappa$ and $\gamma$ given $\theta_{1:M}$, allowing hyperparameter terms to be dropped. Recall $P(o_N|\theta_m,c=m) = \Dir(o_N;\theta_{m}\vec{1}_{m}+\vec{1})$, the Dirichlet density at $o_N$. Thus, \cref{eq:theta_c_inference_orig} provides a generative distribution for low-level observations (after noise parameter inference), and \cref{eq:classification_update_original} provides a recursive filtering rule for macro-observations given each new observation $o_N$. Combined, these equations provide a macro-observation model and filtering scheme which can be used in Dec-POSMDP search algorithms.



\begin{figure}[t]
	\centering
	\hfill
	\includegraphics[trim={0 1cm 0 0.4cm},width=0.5\textwidth]{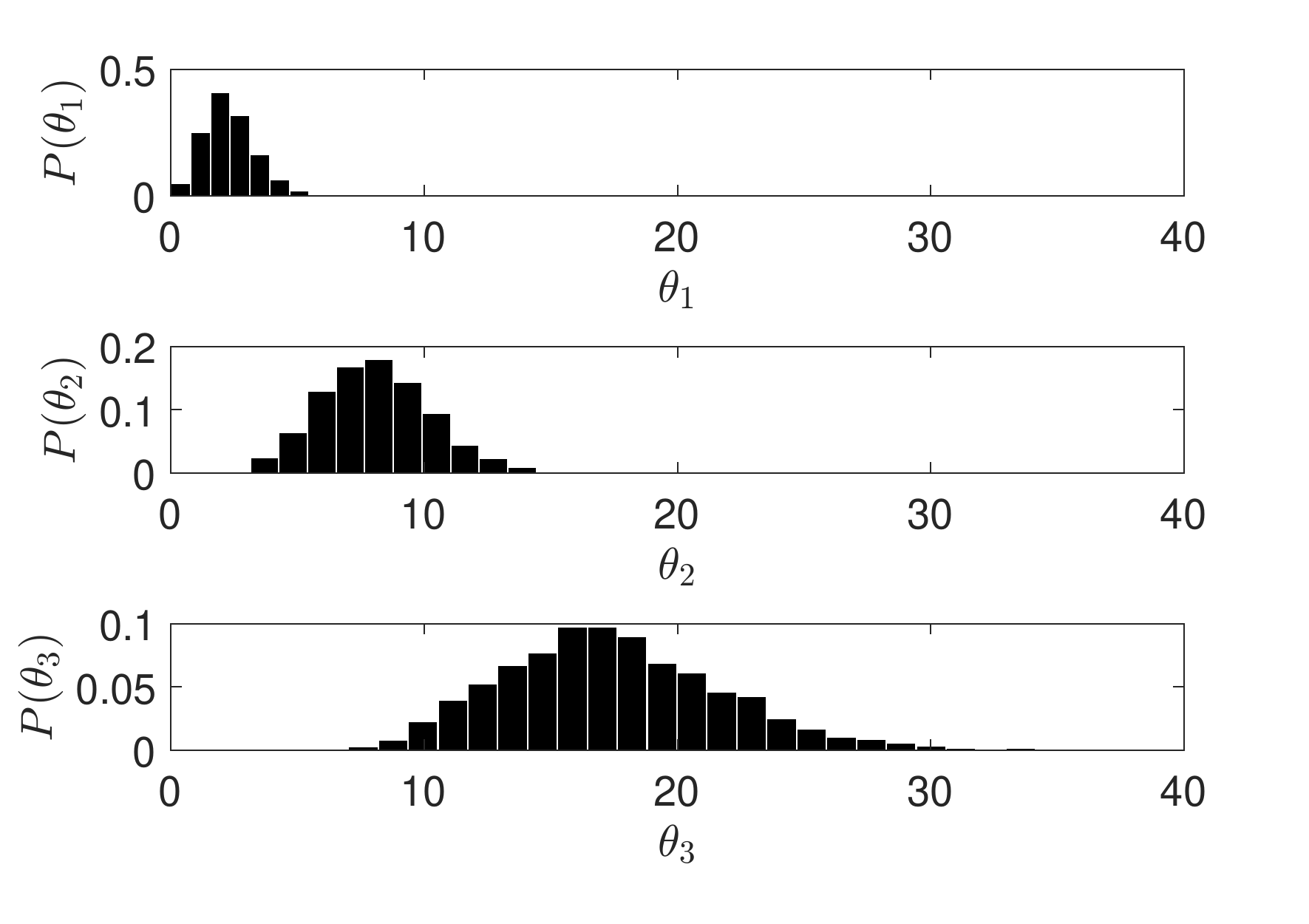}
	\caption{Inferred noise parameters $\theta$ for the $M=3$ classification problem illustrated in \cref{fig:example_observation_models}. True noise parameter values are $\theta_1=1$, $\theta_2=6$, $\theta_3=20$. }
	\label{fig:theta_posterior}
\end{figure}

\begin{figure}[t]
	\vspace{-12pt}
	\centering
	\begin{subfigure}[t]{0.22\textwidth}
		\centering
		\includegraphics[clip,trim={0 0.5cm 0 0.5cm},width=1\textwidth]{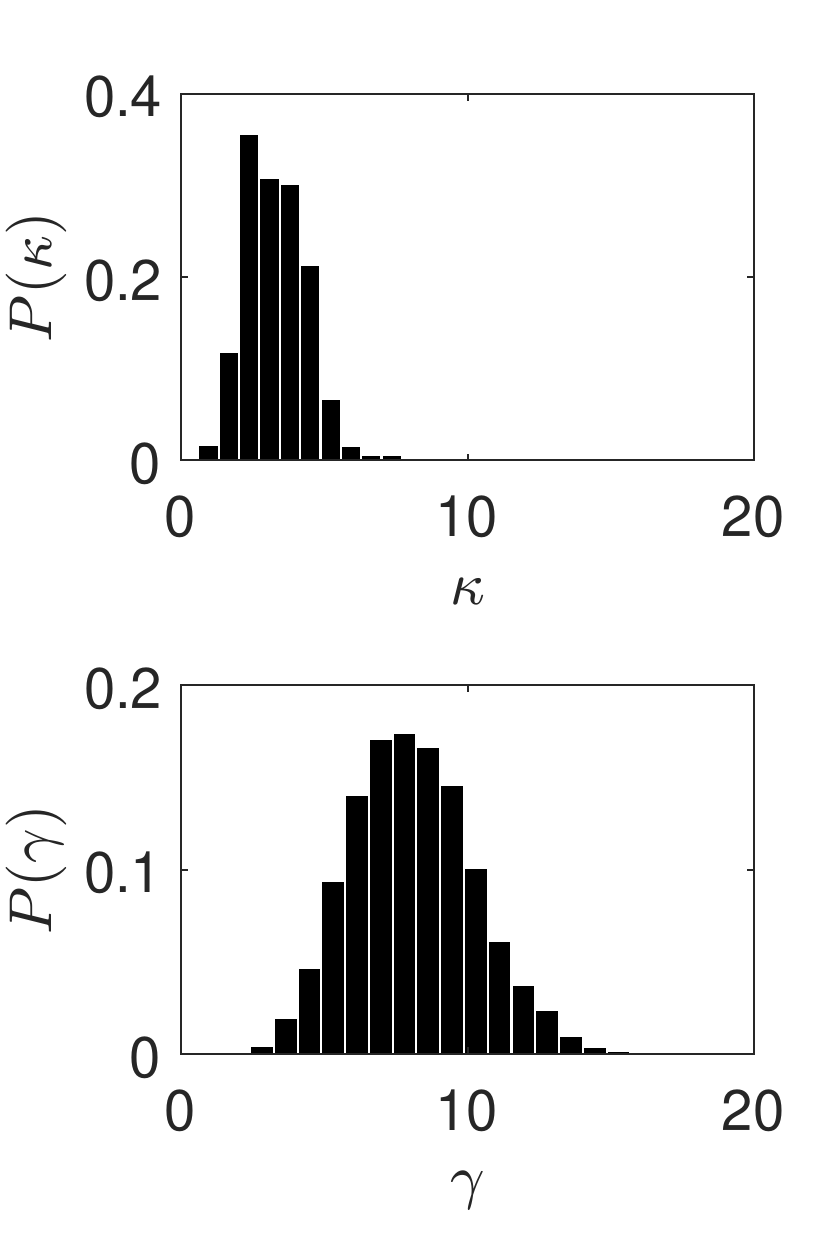}
		\caption{Posterior distributions of $\kappa$ and $\gamma$.}\label{fig:kg_posterior}
	\end{subfigure}
	\hfill
	\begin{subfigure}[t]{0.22\textwidth}
		\centering
		\includegraphics[clip,trim={0 0.5cm 0 0.5cm}, width=1\textwidth]{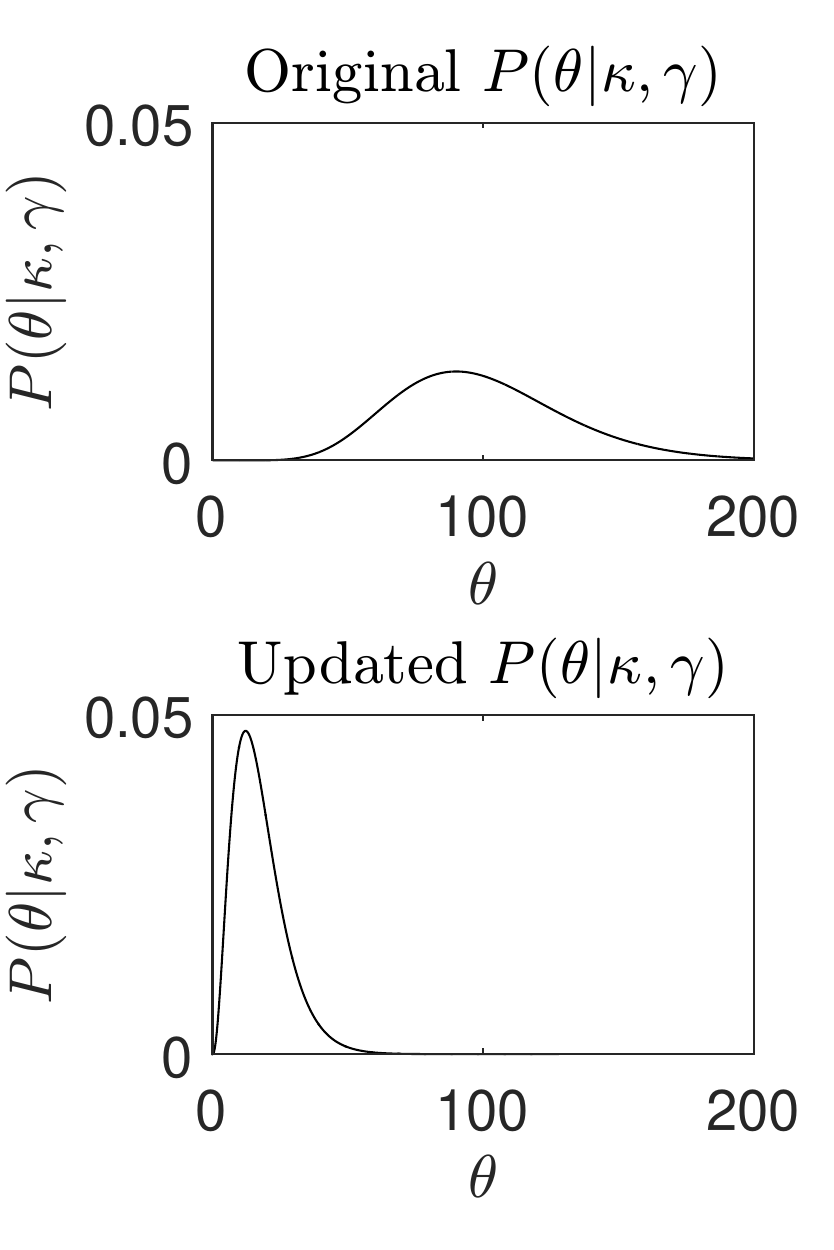}
		\caption{$\theta$ prior before/after hyperparameter update.}
		\label{fig:fhandle_theta_prior_posterior}
	\end{subfigure}
	\caption{Inference of high-level noise parameters $\kappa$, $\gamma$. Median hyperparameters were used for the plots on the right.}
	\label{fig:kg_theta_posteriors}
\end{figure}

To summarize, the proposed HBNI approach uses the collection of classification observations $o_i$ to calculate a posterior distribution on noise parameters $\theta_{1:M}$ for each object class, and shared hyperparameters $\kappa$ and $\gamma$. These noise distributions are then used for online streaming of class probability macro-observations. While HBNI noise inference is computationally efficient and can be conducted online, the complexity of Dec-POSMDPs means that existing sampling-based policy search algorithms are run offline. Thus, integration of HBNI macro-observations into Dec-POSMDPs is a three-fold process. First, domain data is collected and HBNI noise inference of parameters and hyperparameters is conducted, resulting in a generative observation distribution. This distribution is then used for domain sampling and policy search in Dec-POSMDP search algorithms. The resulting policy is then executed online, with HBNI-based filtering used to output macro-observations. The generative nature of HBNI allows usage of complex, durative macro-observation processes, which can filter the stream and output a macro-observation \emph{only when} a desired confidence level is reached.

\section{Simulated Experiments}
This section validates HBNI's performance in comparison to noise-agnostic filtering schemes, before integration into Dec-POSMDPs. As stated earlier, an MCMC approach is used to compute the posterior over $\theta_{1:M}$, $\kappa$, and $\gamma$. Specifically, the experiments conducted use a Metropolis-Hastings (MH) \cite{hastings70} sampler with an asymmetric categorical proposal distribution for underlying classes $c_i$, with high weight on previously-proposed class and low weight on remaining classes (given uniform random initialization). Gaussian MH proposals are used for transformed variables $\log(\theta_m)$, $\log(\kappa)$, and $\log(\gamma)$.

\cref{fig:theta_posterior} shows noise parameter ($\theta_{1:M}$) posterior distributions for the $M=3$ problem outlined in \cref{fig:example_observation_models}. Parameter inference was conducted using only $N=15$ classification observations $o_i$ (5 from each class). Despite the very limited number of observations, the posterior distributions provide reasonable inferences of the true underlying noise parameters. 

Hyperparameter ($\kappa$, $\gamma$) posteriors are shown in \cref{fig:kg_posterior}. Recall these shared parameters capture trends in outputs $o_i$ which indicate shifts in classification confidence levels (for all classes) due to domain-level uncertainty. To test sensitivity of $\theta_m$ inference to the hyperparameters, priors for $\kappa$ and $\gamma$ were chosen such that (on average) they indicate very high values of $\theta_m$ (\cref{fig:fhandle_theta_prior_posterior}, top). This sets a prior expectation of near-perfect outputs from classifiers (median $\theta_m=100$). However, given only $N=15$ classifier observations, posteriors of $\kappa$ and $\gamma$ shift to indicate much lower overall classification confidence $\theta_m$ (\cref{fig:fhandle_theta_prior_posterior}, bottom). $P(\theta_m|\kappa,\gamma)$ has now shifted to better capture the range of noise parameters expected in the domain. This sharing of high-level noise statistics improves filtering of subsequent observations (even if from an entirely new class).

\begin{figure}[t]
	\centering
	\includegraphics[trim={0 0.3cm 0 0.0cm},width=0.49\textwidth]{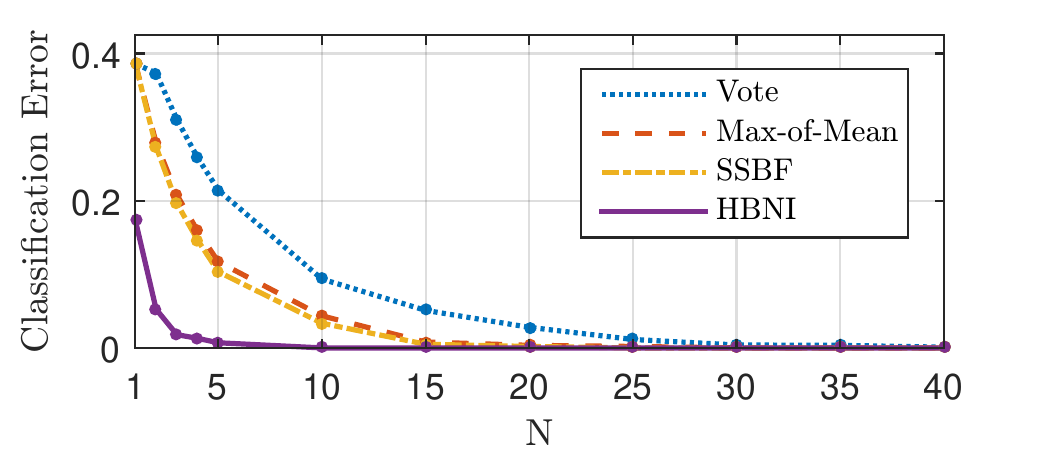}
	\caption{Filtering error for varying observation lengths $N$.}
	\label{fig:fhandle_method_comparison}
	\vspace{-3pt}
\end{figure}

HBNI classification error is evaluated against the voting, max-of-mean, and SSBF methods discussed in \cref{subsec:classification_filtering_problem}. \cref{fig:fhandle_method_comparison} shows results for varying number $N$ of class observations $o_i$, with 2000 trials used to calculate error for each case. Voting performs poorly as it disregards class probabilities altogether. HBNI significantly outperforms the other methods, requiring 5-10 observations to converge to the true object class for all trials. The other methods need 4-5 times the number of observations to match HBNI's performance. One interesting result is that for $N=1$, predictions for voting, max-of-mean, and SSBF are equivalent. However, due to noise modeling, HBNI makes an informed decision regarding underlying class, leading to lower classification error. 

\section{Hardware Experiments}
This section evaluates HBNI on a robotics platform to ascertain the benefits of noise modeling in real-world settings. It then showcases multi-robot Dec-POSMDP decision-making in hardware using HBNI-based macro-observations. 

\subsection{Underlying (Low-Level) Classification Framework}
Low-level classifier training is conducted on a dataset of 3 target vehicle classes (\texttt{iRobot}, \texttt{Quadrotor}, \texttt{Racecar}) in a well-lit room, using a QVGA-resolution webcam (\cref{fig:training_images_overview}). 100 snapshots of each object type are used for training, including crops and mirror images for increased translational and rotational invariance. Feature extraction is done using a Convolutional Neural Net (CNN) implemented in Caffe \cite{jia2014caffe} (though the proposed HBNI approach is agnostic to underlying classifier type). Images are center-cropped with 10\% padding and resized to 227$\times$227 resolution. Features are extracted from the 8-th fully connected layer of an AlexNet \cite{krizhevsky2012imagenet} trained on the ILSVRC2012 dataset \cite{ILSVRC15}. These features are used to train a set of Support Vector Machines (SVMs), with a one-vs-one approach for multi-class classification. 
As SVMs are inherently discriminative classifiers, probabilities $o_i$ for each image $f_i$ are calculated using Platt Scaling, by fitting a sigmoid function to SVM scores \cite{Platt99probabilisticoutputs}. These probabilities are then processed using HBNI-based macro-observations.


\subsection{Hardware Platform}\label{subsec:hardware_platform}
\begin{figure}[t]
	\centering
	\includegraphics[trim={0 0.3cm 0 -0.5cm},width=0.4\textwidth]{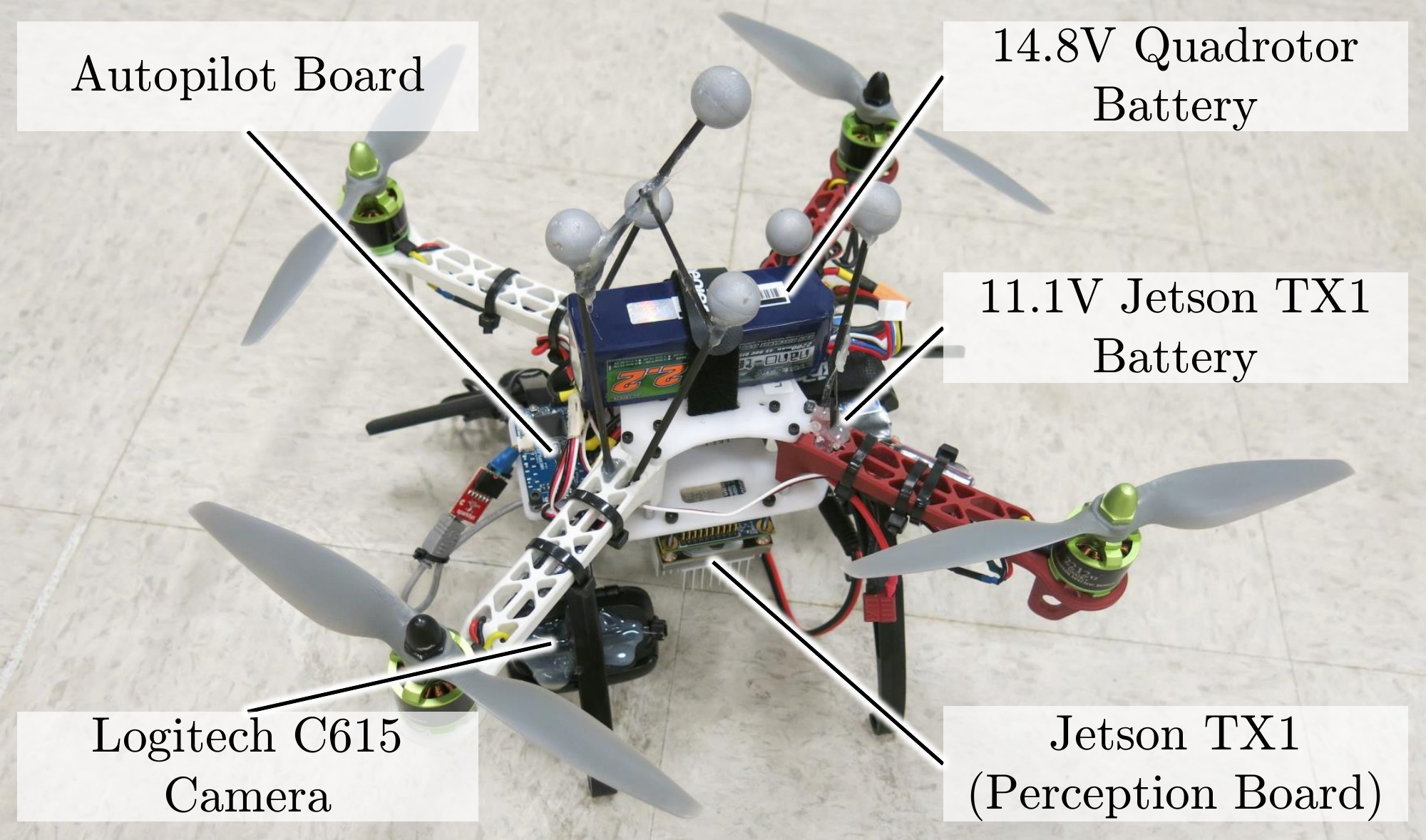}
	\vspace{-2pt}
	\caption{Hardware overview.}
	\label{fig:jetson_quad_overview}
	\vspace{-12pt}
\end{figure}

DJI F330 quadrotors with custom autopilots are used for the majority of experiments (\cref{fig:jetson_quad_overview}), with a Logitech C615 webcam for image capture. The macro-observation pipeline is executed on an onboard NVIDIA Jetson TX1, powered using a dedicated 3-cell 1350mAh LiPo battery. Runtime for the underlying classifier is 49$\pm$5ms per frame, and the entire pipeline (including communication and filtering) executes fully onboard at approximately 20 frames per second.

\subsection{Results: HBNI-based Macro-Observations}
\begin{figure}[t]
	\centering
	\begin{subfigure}[t]{0.115\textwidth}
		\centering
		\includegraphics[width=0.9\textwidth]{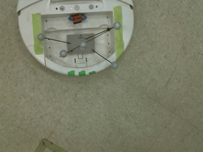}
		\caption{\centering\texttt{iRobot} class example.}
		\label{fig:training_images_overview_v2_irobot}
	\end{subfigure}
	\hfill
	\begin{subfigure}[t]{0.115\textwidth}
		\centering
		\includegraphics[width=0.9\textwidth]{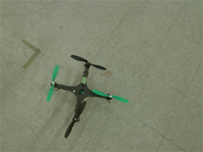}
		\caption{\centering\texttt{Quadrotor} class example.}
		\label{fig:training_images_overview_v2_quad}
	\end{subfigure}
	\hfill
	\begin{subfigure}[t]{0.115\textwidth}
		\centering
		\includegraphics[width=0.9\textwidth]{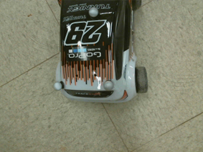}
		\caption{\centering\texttt{Racecar} class example.}
		\label{fig:training_images_overview_v2_racecar}
	\end{subfigure}
	\hfill
	\begin{subfigure}[t]{0.115\textwidth}
		\centering
		\includegraphics[width=0.9\textwidth]{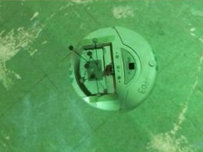}
		\caption{\centering Test domain conditions.}
		\label{fig:training_images_overview_v2_proj_on}
	\end{subfigure}
	\vspace{-3pt}
	\caption[]{Comparison of training and test images in domains with varying lighting conditions.}
	\label{fig:training_images_overview}
	\vspace{-9pt}
\end{figure}

Classification robustness is verified using an augmented reality testbed \cite{Omidshafiei16_CSM_short} to change domain lighting conditions. In contrast to the well-lit images used to train the underlying classifier (\cref{fig:training_images_overview_v2_irobot,fig:training_images_overview_v2_quad,fig:training_images_overview_v2_racecar}), test images have textured backgrounds and dim lighting which reduce camera shutter speed, increasing blur (\cref{fig:training_images_overview_v2_proj_on}). Experiments are designed to simulate typical scenarios in robotics where the training dataset is not fully representative of mission test data.

\begin{figure}[h!]
	\centering
	\begin{subfigure}[t]{0.45\textwidth}
		\centering
		\includegraphics[width=1\textwidth]{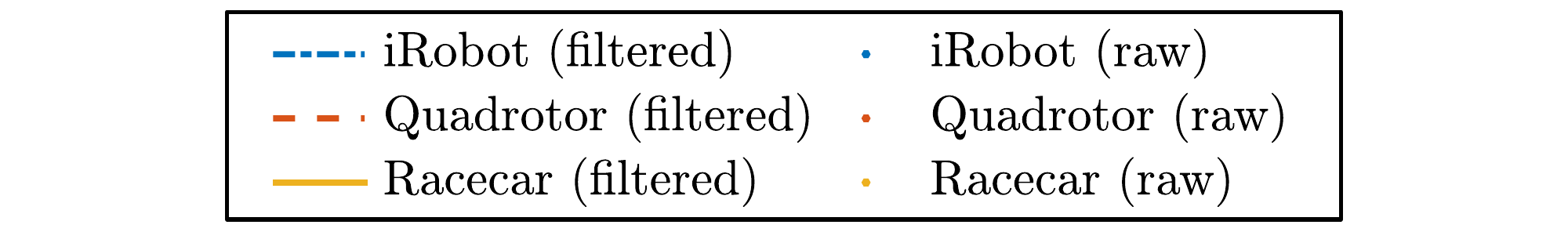}
	\end{subfigure}
	\\
	\begin{subfigure}[t]{0.235\textwidth}
		\centering
		\includegraphics[trim={0 0.3cm 0 0},width=1\textwidth]{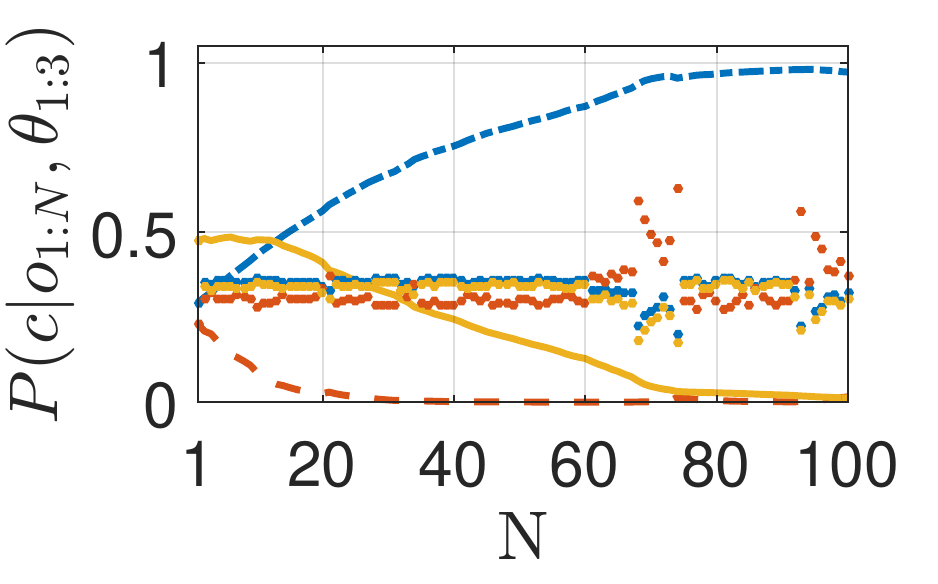}
		\caption{SSBF posterior over time.}
		\label{fig:method_comparison_actual_data_all_data_true_class_1_SSBF}
	\end{subfigure}
	\hfill
	\begin{subfigure}[t]{0.235\textwidth}
		\centering
		\includegraphics[trim={0 0.3cm 0 0},width=1\textwidth]{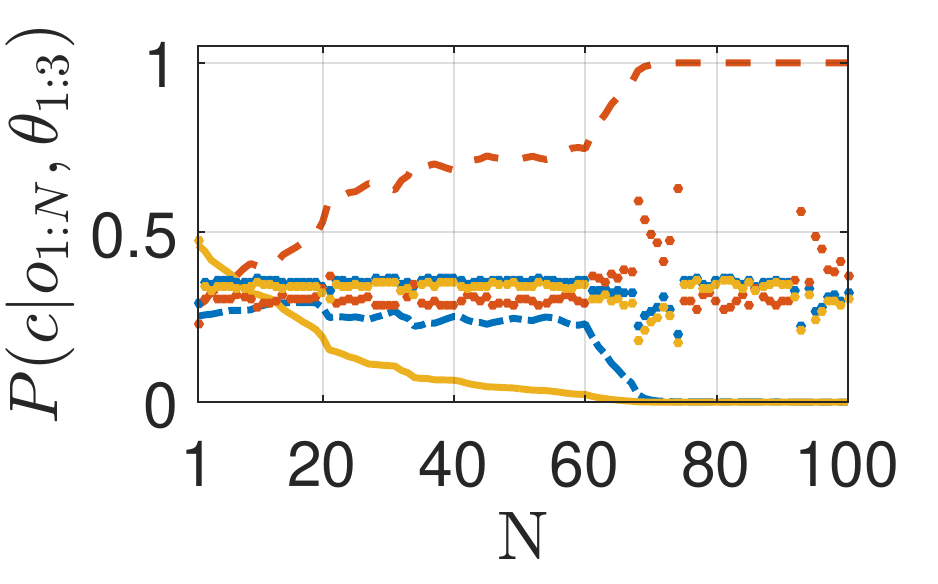}
		\caption{HBNI posterior over time.}
		\label{fig:method_comparison_actual_data_all_data_true_class_1_HBNI}
	\end{subfigure}
	\vspace{-3pt}
	\caption{Comparison of SSBF and HBNI filtering, recorded on a moving quadrotor. True object class is \texttt{Quadrotor}.} 
	\label{fig:method_comparison_hardware}
\end{figure}

\begin{figure*}[t]
	\centering
	\includegraphics[trim={0 0.2cm 0 -0.2cm},width=1\textwidth]{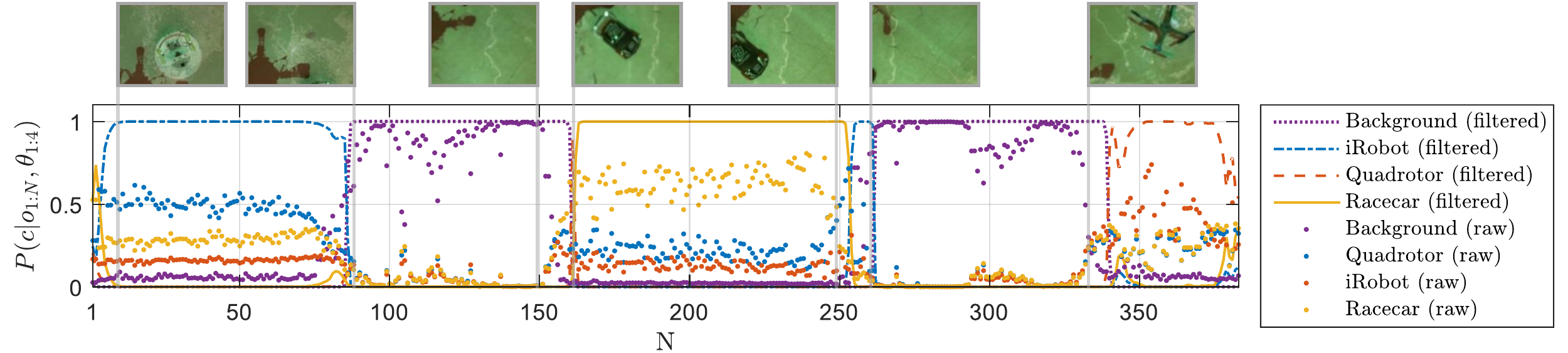}
	\caption{HBNI-based filtered macro-observations onboard a moving quadrotor (example frames indicated).}
	\label{fig:fhandle_rss_python_plots}
	\vspace{-19pt}
\end{figure*}

Filtered classification results for the test dataset are shown in \cref{fig:method_comparison_hardware}. In new lighting conditions, classification of the \texttt{Quadrotor} object class is particularly difficult, resulting in nearly equal raw probabilities $o_i$ amongst all three classes (raw data in \cref{fig:method_comparison_hardware}). Noise-agnostic filters such as SSBF fail to correctly classify the object as a \texttt{Quadrotor}, instead classifying it as an \texttt{iRobot} with high confidence (filtered output in \cref{fig:method_comparison_actual_data_all_data_true_class_1_SSBF}). Moreover, probability of the \texttt{Quadrotor} class asymptotically approaches zero as more observations are made. In contrast, HBNI infers underlying noise, leading to robust classification of the \texttt{Quadrotor} object after only 7 frames (\cref{fig:method_comparison_actual_data_all_data_true_class_1_HBNI}). In the $N=70$ to $N=75$ range, due to improved lighting, raw classifier probabilities increase for the \texttt{Quadrotor} class. SSBF only slightly lowers its probability of the object being an \texttt{iRobot}, whereas the HBNI approach significantly increases probability of the true \texttt{Quadrotor} class. \cref{fig:fhandle_rss_python_plots} shows HBNI macro-observations on a quadrotor exploring an environment with multiple objects. The results indicate that HBNI accurately classifies objects onboard a moving robot in noisy domains. For additional HBNI results and analysis, readers can refer to our technical report \cite{Omidshafiei16_HBNIarxiv}.


\subsection{Results: Multi-Robot Decision-Making}

\begin{figure}[t]
	\includegraphics[width=1\linewidth]{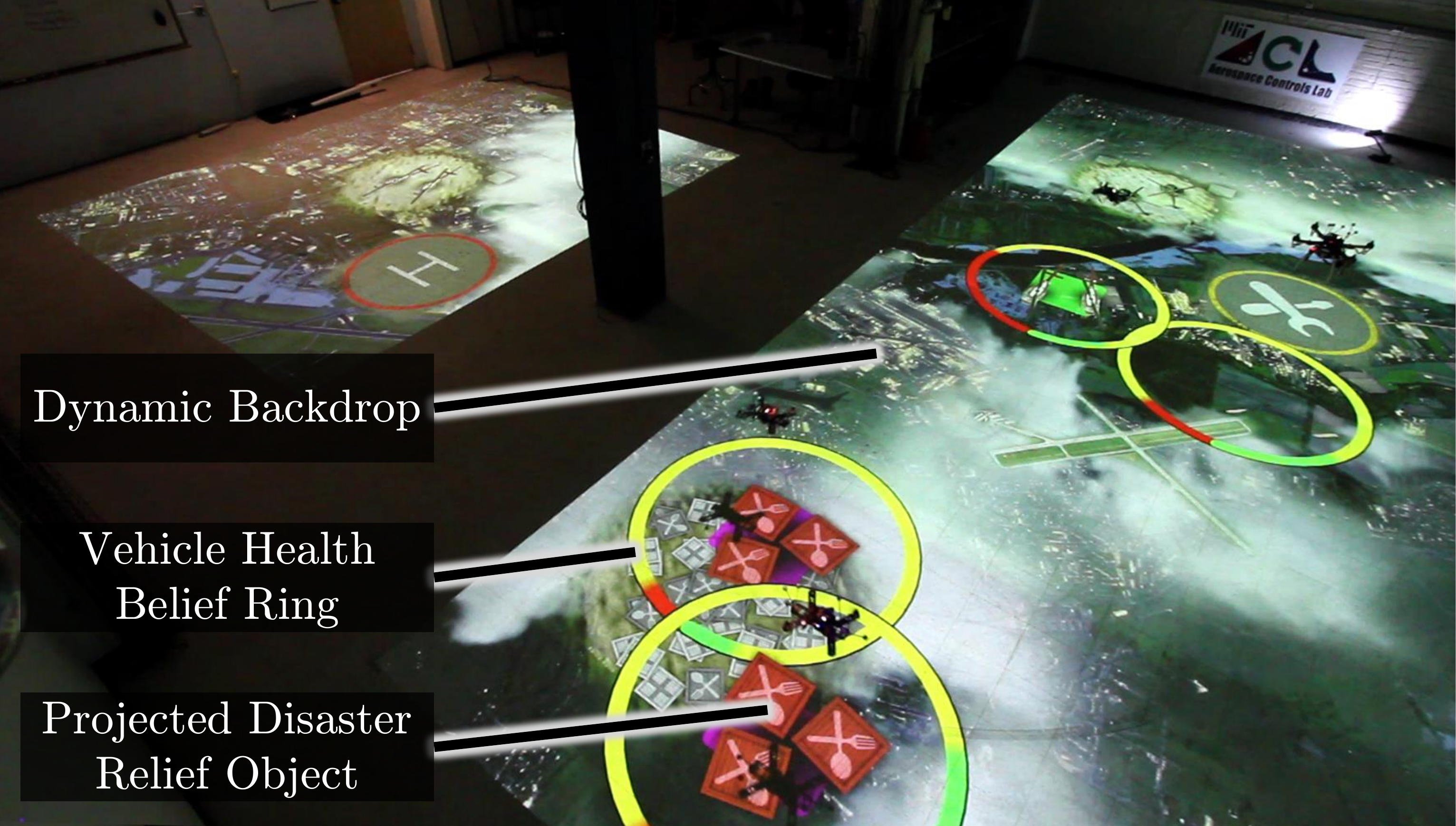}
	\caption{Health-aware multi-quadrotor disaster relief domain, via macro-observation-based planning in dynamic environment. Video: \url{https://youtu.be/XXYSAmdHn38}.}
	\label{fig:boeing_demo_2016_general}
\end{figure}

HBNI-based macro-observations were integrated into the Dec-POSMDP framework (as described in \cref{sec:macro_observations}) and evaluated on a multi-robot health-aware disaster relief domain (\cref{fig:boeing_demo_2016_general}). This is an extension of the Dec-POSMDP package delivery domain \cite{Omidshafiei15_ICRA} involving a team of quadrotors. Disaster relief objects of 6 types (\texttt{ambulance}, \texttt{police\_car}, \texttt{medical\_copter}, \texttt{news\_copter}, \texttt{food\_crate}, \texttt{medical\_crate}) are randomly generated at 2 bases, each with an associated delivery destination (\texttt{hospital}, \texttt{airport}, or \texttt{crate\_zone}). Nine MAs are available for execution by each robot: \emph{Go to $Base_{1}/Base_{2}/Hospital/Airport/Crate\_zone$}, \emph{Go to repair station for maintenance}, \emph{Infer object class with 95\% confidence}, \emph{Pick up disaster relief object}, \emph{Put down disaster relief object}. Quadrotors are outfitted with the hardware discussed in \cref{subsec:hardware_platform} and use HBNI to infer the underlying disaster relief object class during policy execution. The team receives a reward for each object delivered to the correct destination. Quadrotors also receive noisy observations from onboard health sensors and maintain a belief distribution over their underlying health state (high, medium, and low health), indicated by colored rings in \cref{fig:boeing_demo_2016_general}. Robots with low health take longer to complete MAs, thereby reducing overall team reward due to the discount factor in \cref{eq:decposmdp_eval}. Perception data is collected and used to train the HBNI-based macro-observation process, which is then used for Dec-POSMDP policy search via the Graph-based Direct Cross Entropy algorithm \cite{Omidshafiei16_ICRA}. 


MAs in this domain have probabilistic success rates and completion times. An augmented reality system is used to display bases, disaster relief objects, and delivery destinations in real-time in the domain. The domain includes shadows and camera noise, but perception uncertainty is further increased by projecting a dynamic day-night cycle and moving backdrop of clouds on the domain. 

Our video attachment shows this multi-robot mission executed on a team of quadrotors. HBNI inference occurs onboard, with the necessary number of low-level observations processed to achieve high confidence. Mission performance matches that of previous (simpler) results for this domain which simulated all observations \cite{Omidshafiei16_ICRA}. To the best of our knowledge, this is the first demonstration of real-time, CNN-based classification running onboard quadrotors in a team setting. It is also the first demonstration of data-driven multi-robot semantic-level decision-making using Dec-POSMDPs.


\section{Conclusion}
This paper presented a formalization of macro-observation processes used within Dec-POSMDPs, targeting scalability improvements for real-world robotics. A hierarchical Bayesian approach was used to model semantic-level macro-observations. This approach, HBNI, infers underlying noise distributions to increase classification accuracy, resulting in a generative macro-observation model. This is especially useful in robotics, where perception sensors are notoriously noisy. The approach was demonstrated in real-time on moving quadrotors, with classification and filtering performed onboard at approximately 20 frames per second. The novel macro-observation process was then integrated into a Dec-POSMDP and demonstrated in a probabilistic multi-robot health-aware disaster relief domain. Future work includes extension of existing Dec-POSMDP algorithms to online settings to leverage the computational-efficiency of HBNI. 


\bibliographystyle{IEEEtran}
\bibliography{TemplateFiles/shayegan_bib,BIB_all/ACL_all,BIB_all/ACL_bef2000,BIB_all/ACL_Publications,references}

\end{document}